\documentclass[apjl]{emulateapj}
\slugcomment{(To appear in the ApJ Letters)}
\newcommand{\um}{\mbox{$\,\mu{\rm m}$}}

\newcommand{\etal}{et al.~}

\newcommand{\Herschel}{{\it Herschel}}

\newcommand{\NeII}{[Ne\,{\sc ii}]}
\newcommand{\NeV}{[Ne\,{\sc v}]}
\newcommand{\LIR}{\mbox{$L_{\rm IR}$}}
\newcommand{\LFIR}{\mbox{$L_{\rm FIR}$}}

\newcommand{\NII}{\mbox{[N\,{\sc ii}]}}
\newcommand{\CII}{\mbox{[C\,{\sc ii}]}}
\newcommand{\NIIline}{\mbox{[N\,{\sc ii}]$_{205\mu {\rm m}}$}}

\begin{document}

\shorttitle{CO\,(7$-$6) and \NII\ 205\um\ Line Emission in Galaxies}
\shortauthors{Lu et al.}

\title{Measuring Star-formation Rate and Far-Infrared Color in High-redshift Galaxies \\ 
       Using the CO\,(7$-$6) and \NII\ 205\um\ Lines\footnotemark[$\star$]}

\author{Nanyao Lu\altaffilmark{1}, 
Yinghe Zhao\altaffilmark{1, 2, 3}, 
C. Kevin Xu\altaffilmark{1}, 
Yu Gao\altaffilmark{2, 3}, 
Tanio D\'iaz-Santos\altaffilmark{4,   5}, 
Vassilis Charmandaris\altaffilmark{6,7,8}, \\
Hanae Inami\altaffilmark{9},
Justin Howell\altaffilmark{1}, 
Lijie Liu\altaffilmark{2, 10}, 
Lee Armus\altaffilmark{4}, 
Joseph M. Mazzarella\altaffilmark{1},
George C. Privon\altaffilmark{11}, \\
Steven D. Lord\altaffilmark{12},
David B. Sanders\altaffilmark{13},
Bernhard Schulz\altaffilmark{1}, and
Paul P. van der Werf\altaffilmark{14}
}
\altaffiltext{1}{Infrared Processing and Analysis Center, California Institute of Technology, MS 100-22, Pasadena, CA 91125, USA; lu@ipac.caltech.edu}
\altaffiltext{2}{Purple Mountain Observatory, Chinese Academy of Sciences, Nanjing 210008, China}
\altaffiltext{3}{Key Laboratory of Radio Astronomy, Chinese Academy of Sciences, Nanjing 210008, China}
\altaffiltext{4}{Spitzer Science Center, California Institute of Technology, MS 220-6, Pasadena, CA 91125, USA}
\altaffiltext{5}{Nucleo de Astronomia de la Facultad de Ingenieria, Universidad Diego Portales, 
                 Av. Ejercito Libertador 441, Santiago, Chile}
\altaffiltext{6}{Department of Physics, University of Crete, GR-71003 Heraklion, Greece}
\altaffiltext{7}{IAASARS, National Observatory of Athens, GR-15236, Penteli, Greece}
\altaffiltext{8}{Chercheur Associ\'e, Observatoire de Paris, F-75014, Paris, France}
\altaffiltext{9}{National Optical Astronomy Observatory, Tucson, AZ 85719, USA}
\altaffiltext{10}{Max-Planck-Institut f\"ur Radioastronomie, Auf dem H\"ugel 69, D-53121 Bonn, Germany}
\altaffiltext{11}{Departamento de Astronom\'ia, Universidad de Concepci\'on, Casilla 160-C, Concepci\'on, Chile}
\altaffiltext{12}{The SETI Institute, 189 Bernardo Ave, Suite 100, Mountain View, CA 94043, USA}
\altaffiltext{13}{University of Hawaii, Institute for Astronomy, 2680 Woodlawn Drive, Honolulu, HI 96822, USA}
\altaffiltext{14}{Leiden Observatory, Leiden University, PO Box 9513, 2300 RA Leiden, The Netherlands}
%
\footnotetext[$\star$]{
Based on \Herschel\ observations. 
\Herschel\ is an ESA space observatory with science 
instruments provided by European-led Principal Investigator consortia 
and with important participation from NASA.}


\begin{abstract}
To better characterize the global star formation (SF) activity in a 
galaxy, one needs to know not only the star formation rate (SFR) but 
also the rest-frame, far-infrared (FIR) color (e.g., the 60-to-100\um\ 
color, $C(60/100)$] of the dust emission.  The latter probes the average 
intensity of the dust heating radiation field and scales statistically 
with the effective SFR surface density in star-forming galaxies including
(ultra-)luminous infrared galaxies [(U)LIRGs].
To this end, we exploit here a new spectroscopic approach involving 
only two emission lines: CO\,(7$-$6) at 372\um\ and \NII\ at 205\um\ 
(\NIIline).  For local (U)LIRGs, the ratios of the CO\,(7$-$6) luminosity
($L_{\rm CO\,(7-6)}$) to the total infrared luminosity 
(\LIR; 8$-$1000\um) are fairly tightly distributed (to within $\sim$0.12
dex) and show {\it little} dependence on $C(60/100)$.  This makes 
$L_{\rm CO\,(7-6)}$ a good SFR tracer, which is less contaminated by 
active galactic nuclei (AGN) than \LIR\ and may also be much less sensitive
to metallicity than $L_{\rm CO\,(1-0)}$.  Furthermore,
the logarithmic \NIIline/CO\,(7$-$6) luminosity ratio is fairly steeply 
(at a slope of $\sim$$-1.4$) correlated with $C(60/100)$, with a modest
scatter ($\sim$0.23 dex).  This makes it a useful estimator on $C(60/100)$
with an implied uncertainty of $\sim$0.15 [or $\lesssim$4\,K in the dust 
temperature ($T_{\rm dust}$) in the case of a graybody emission with $T_{\rm dust} 
\gtrsim 30$\,K and a dust emissivity index $\beta \ge 1$].  
Our locally calibrated SFR and 
$C(60/100)$ estimators are shown to be consistent with the published 
data of (U)LIRGs of $z$ up to $\sim$6.5. 
\end{abstract}

\section{INTRODUCTION} \label{sec1}

\setcounter{footnote}{10}

Luminous infrared galaxies (LIRGs: with an 8-1000\um\ $L_{\rm IR} > 
10^{11}L_{\sun}$; Sanders \& Mirabel 1996), including ultra-luminous
ones (ULIRGs, $L_{\rm IR} > 10^{12}L_{\sun}$), dominate the cosmic 
star formation (SF) at $z \gtrsim 1$ (e.g., Le Fl\'och \etal 2005).
For $z \sim 1$ to 3, these galaxies are mixtures of two populations 
based on their prevalent ``SF mode'': (a) mergers dominated by 
nuclear starburst with warm far-infrared (FIR) colors and a high SF 
efficiency (SFE) similar to that in local ULIRGs, and (b) gas-rich 
disk galaxies with disk SF and SFE comparable to
local spirals (e.g., Daddi et al.~2010);
more ULIRGs belong to the latter ``main-sequence'' (MS) population
(e.g., Elbaz et al.~2011).
However, the current perception that the typical spectral energy 
distribution (SED) of the dust emission in the high-$z$, MS galaxy 
population is merely a ``scaled-up'' SED of local normal galaxies
remains unproven: if the size of the effective SF region 
in a galaxy is fixed, an increasing \LIR\ implies a higher effective 
SFR surface density ($\Sigma_{\rm SFR}$), which is known to lead 
to a warmer 60-to-100\um\ color, $C(60/100)$, (thus, the shape
of the SED) for both normal galaxies and (U)LIRGs (Chanial \etal 2007;
Liu \etal 2015; also see \S3.2).
As demonstrated by Rujopakarn et al.~(2011), the high-$z$ (U)LIRGs
from the MS population are comparable in size to the local star-forming
galaxies, but with a much higher $\Sigma_{\rm SFR}$.
In general, $C(60/100)$ probes the average intensity of the dust heating 
radiation field (e.g., Draine \& Li 2007), and both SFR and
$C(60/100)$ should be measured in order to more fully characterize 
the SF activity in galaxies.

The conventional way to do so is to obtain a full dust SED from 
which both the SFR (from \LIR) and $C(60/100)$ can be deduced.  
For high-$z$ galaxies, this usually requires multiple photometric 
measurements covering a wide wavelength range, as illustrated 
in the recent studies on 3 galaxies at $z \sim 5$-6 (Riechers 
\etal 2013; Gilli et al.~2014; Rawle et al.~2014).  Furthermore, 
as $z$ increases, accurate continuum photometry becomes 
tougher due to relatively bright background.  A promising alternative 
is to measure SFR and $C(60/100)$ using spectral lines
in the FIR/sub-mm.  A recent spectroscopic survey with 
the {\it Herschel Space Observatory} (\Herschel) on a large sample
of LIRGs from the Great Observatories All-Sky LIRG 
Survey (GOALS; Armus et al.~2009) revealed a remarkable one-to-one
relation between the luminosity summed over the CO rotational 
transitions in the mid-$J$ regime ($5 \le J \lesssim 10$) and \LIR\ 
(Lu \etal 2014; hereafter Paper I). Here we exploit 
the scenario of using only 
the CO\,(7$-$6) line luminosity, $L_{\rm CO\,(7-6)}$, as a SFR
tracer.  
Furthermore, we show that, for the local (U)LIRGs, the \NII\ 205\um\ 
line (hereafter simply as \NII) to CO\,(7$-$6) flux 
ratio is fairly steeply correlated with $C(60/100)$ with a 
modest scatter. As a result, it can serve as a useful estimator
on $C(60/100)$.

In the remainder of this Letter, we describe the galaxy samples 
and data used in \S2,  present our analysis and results in \S3, 
and compare our results to the existing observations of distant 
galaxies in \S4.
 
\section{Data Samples} \label{sec2}

\subsection {Local LIRGs}
Paper I described a \Herschel\ spectroscopic 
survey of a flux-limited set of 125 LIRGs from GOALS using the Spectral
and Photometric Imaging REceiver (SPIRE; Griffin et al.~2010). 
While the detailed data will be presented elsewhere (Lu et al., 
in preparation), the measured CO and \NII\ fluxes
based on the point-source flux calibration, as described in Paper I
and Zhao \etal (2013), are used here.  
The fluxes of the 
\CII\ line at 158\um\ (hereafter as \CII) for our galaxies
were taken from D\'iaz-Santos \etal (2013), obtained with the \Herschel\ 
Photodetector Array Camera and Spectrometer (PACS; Poglitsch 
et al.~2010).  These are also point-source calibrated fluxes, 
approaching the total flux for sources with a \CII\ extent not 
too extended relative to the PACS beam of $\sim$12\arcsec (full 
width at half maximum or FWHM).

For each GOALS galaxy, the \NeV\ 14.3\um\ to \NeII\ 12.8\um\ 
line ratio (hereafter \NeV/\NeII) or its upper limit is available
in Inami et al.~(2013).  Five galaxies in our GOALS/SPIRE sample
have \NeV/\NeII\ $>0.65$, for which the AGN 
contribution to the total bolometric luminosity is likely greater 
than 50\% (Farrah \etal 2007).

\subsection {Local ULIRGs}
Containing only 7 ULIRGs, our GOALS/SPIRE sample covers mainly LIRGs, particularly 
lower luminosity ones.  We also obtained 
from the \Herschel\ archive and reduced in the same way the SPIRE 
spectroscopic observations
of 28 ULIRGs (i.e., the local ULIRG sample), which extend our \LIR\ coverage 
to $\sim$10$^{13}\,L_{\odot}$.  These observations were performed in 
the program ``OT1$_-$dfarrah$_-$1'' (PI: D. Farrah).  For many of these galaxies,
the \CII\ fluxes are available from Farrah \etal (2013).  Only one galaxy in 
this sample has \NeV/\NeII\ $>0.65$ (Farrah \etal 2007).

\subsection {Local Dwarf Galaxies}
Our GOALS/SPIRE sample also includes one blue compact dwarf, Haro 11, with a 
metallicity $Z \approx 0.45\,Z_{\odot}$, where $Z_{\odot} = (12 + \log\,{\rm O/H})_{\rm solar}
= 8.7$ (Asplund et al. 2009).  
We obtained archival SPIRE spectra on 3 additional dwarfs: NGC\,4214 ($Z \sim$ 0.36\,$Z_{\odot}$; obsid $=$ 
1342256082; Madden \etal 2013), IC\,10 ($\sim$0.29\,$Z_{\odot}$;
1342246982) and He\,2-10 ($\sim$0.54\,$Z_{\odot}$; 1342245083) 
(PI: V. Lebouteiller).  The metallicity values were taken from R\'emy-Ruyer 
\etal (2013).  Both IC\,10 and NGC\,4214 are extended and their SPIRE 
observations were pointed at the brightest \ion{H}{2} region.
We extracted a CO\,(7$-$6) flux from the point-source calibrated spectrum
of the central detector.  The corresponding 
$f_{\nu}(70\um)$ and $f_{\nu}(100\um)$ were derived by convolving 
the SPIRE beam of CO\,(7$-$6) with the corresponding PACS 
images (R\'emy-Ruyer et al.~2013) and used to
calculate the FIR luminosity (\LFIR; Helou et al.~1985) after inferring 
$f_{\nu}(60\um)$ from a matching FIR model SED from Dale \etal (2001).  
He\,2-10 is infrared compact (Bendo et al.~2012).  Its SPIRE 
observation is a map. We therefore extracted from the map a point-source 
spectrum at the location of the peak brightness
using the task ``specPointSourceExtractor'' in the 
\Herschel\ Interactive Processing Environment software (HIPE).
The extracted spectrum was further corrected for an optimized source extent
of 18\arcsec (Gaussian FWHM) using the HIPE semi-extended source correction 
tool (Wu et al.~2013) before extracting our CO\,(7$-$6) and \NII\ fluxes.
Finally, the total \CII\ flux of He\,2-10 was taken from Cormier \etal (2015).

\section{Analysis} \label{sec3}

\subsection{CO\,(7$-$6) as a SFR Tracer} \label{sec3.1}

We used the PACS 70\um\ continuum images of Chu et al.~(in preparation) 
to further select our GOALS galaxies 
that are not too extended with respect to the SPIRE beam, which 
measures (FWHM) 35\arcsec\ and 17\arcsec\ for CO\,(7$-$6)
and \NII, respectively, in order to use the point-source calibrated
fluxes.  For each galaxy, we calculated $f_{70\mu{\rm m}}(\theta)$, 
the fractional 70\um\ flux within a Gaussian beam of FWHM $\theta$.
The CO\,(7$-$6) and \NII\ 
analyses here are further limited to those GOALS galaxies
satisfying $f_{70\mu{\rm m}}(30\arcsec) > 85\%$ (102 galaxies; 
the average value of $f_{70\mu{\rm m}}(30\arcsec) = 97\%$) and 
$f_{70\mu{\rm m}}(17\arcsec) > 70\%$ (98 galaxies; the average value 
$= 89\%$), respectively.   These were chosen so that at least 75\% of 
the GOALS galaxies in the coldest FIR color ($0.45 \lesssim C(60/100) \le 0.6$)
or smallest \LIR\ ($11 
\lesssim \log L_{\rm IR}/L_{\odot} \le 11.3$) bin meet the criterion, 
and that any systematic effect from the possible aperture flux loss 
is significantly smaller than the sample scatter in the flux ratios 
dealt with here.

In Fig.~1 we plot both $L_{\rm CO\,(7-6)}$/\LIR\ and $L_{\rm CO\,(7-6)}$/\LFIR\ 
as a function of $C(60/100)$ for the local galaxies.  
The AGNs with \NeV/\NeII\ $> 0.65$ are further circled.  
For IC\,10 
and NGC\,4214, we could only obtain their \LFIR. As a result, 
they are only shown in Fig.~1b.

\begin{figure}
\centering
\includegraphics[width=.55\textwidth, bb=40 160 592 718]{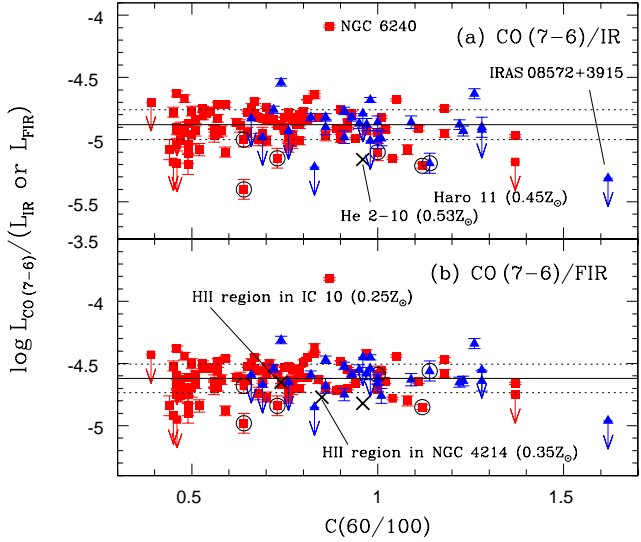}
\vspace{-0.8in}
\caption{
Plots of the logarithmic ratio of $L_{\rm CO\,(7-6)}$ to \LIR\ in (a) 
and to \LFIR\ in (b) against $C(60/100)$, for the 102 selected (see 
the text) GOALS/SPIRE galaxies (red squares) and the local ULIRGs 
(blue triangles).  The error bars shown are at 1\,$\sigma$. 
The non-detections are shown with their 3$\,\sigma$ upper limit.  
The 6 powerful AGNs are further enclosed by a circle.  A few dwarf 
galaxies (large black crosses) are individually labelled.  The solid 
line marks the average [$= -4.88 \pm 0.01$ in (a) or $-4.61 \pm 0.01$ 
in (b)], which also agrees 
well with the median, for the combined (U)LIRG samples and the two 
dashed lines the corresponding sample standard deviation,   
both determined from the detections only (but excluding NGC\,6240 and 
the 6 AGNs).}
\label{Fig1}
\end{figure}

NGC\,6240 is a rare outlier with additional gas heating likely from 
shocks unrelated to the ongoing SF (see Paper I).
The AGNs in Fig.~1 have a lower CO/IR ratio on average.  This is more 
apparent in CO\,(7$-$6)/IR than in CO\,(7$-$6)/FIR, consistent with 
that the lower CO/IR ratio in an AGN is mostly due to the AGN ``contamination'' 
to \LIR (see Paper I).  In this sense, CO\,(7$-$6) is a ``cleaner'' SFR 
tracer than \LIR.  

A galactic CO spectral line energy distribution (SLED) generally 
consists of up to 3 distinct gas components, which dominate the SLED 
at the low (i.e., $J \lesssim 4$), mid ($\sim$5 to $\sim$10) and high $J$ 
($> 10$) regimes, respectively; the mid-$J$ component is directly
related to the ongoing SF (see Paper I; Xu et al.~2015).  On either 
side (in $J$) of this SF-driven component, the ``contamination'' from 
one of the other gas components increases.  We found that CO\,(7$-$6)
traces \LIR\ better than any other mid-$J$ CO line.  For example,
the CO\,(6$-$5)/IR ratios from our sample show a small anti-correlation
with $C(60/100)$, implying a systematic ratio variation of $\sim$0.29 dex
over $0.4 < C(60/100) < 1.3$.

The average CO\,(7$-$6)/IR ratio in Fig.~1 (i.e., the solid line) can be used to 
derive a SFR from $L_{\rm CO(7-6)}$ using the SFR-\LIR\ calibration from
Kennicutt (1998).  Eq.~(1) gives the results, with the quoted uncertainty 
being the sample standard deviation ($\sigma_s \approx 0.12$ dex). Since 
$\sigma_s \approx 0.10$ dex when the total flux of all the mid-$J$ CO lines 
was used (see Paper I), it is $\sim$5\% less accurate when using the CO\,(7$-$6)
line alone to predict \LIR.

\begin{eqnarray}
SFR/(M_{\odot}\,{\rm yr}^{-1})\  & = \ & 1.73\times 10^{-10}\,(L_{\rm IR}/L_{\odot}) \nonumber \\
			         & = \ & 1.31\times 10^{-5.00 \pm 0.12}\,(L_{\rm CO\,(7-6)}/L_{\odot}). \nonumber \\	
				 &
\end{eqnarray}

\begin{figure}
\centering
\includegraphics[width=.55\textwidth, bb=60 150 605 718]{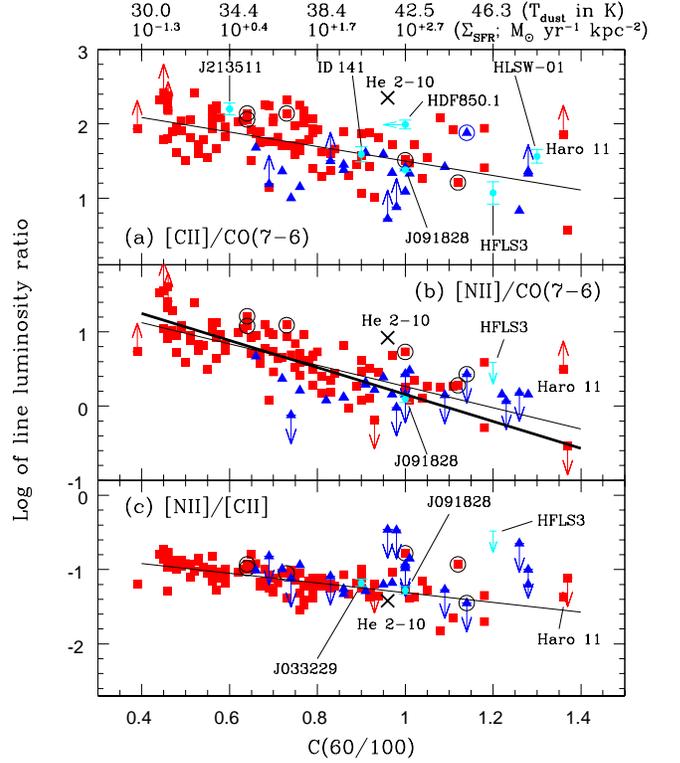}
\caption{
Plots of various line luminosity ratios against $C(60/100)$:  (a) \CII\ 
to CO\,(7$-$6), (b) \NII\ to CO\,(7$-$6), and (c) \NII/\CII, using the same symbols
and color schemes as in Fig.~1.  Only 98 GOALS galaxies are plotted (see the text). 
The error bars were omitted here as they are smaller than the scatter. 
The thin solid line indicates a vertical least-squares fit to all the detections
of the local (U)LIRGs.
The thick solid line in (b) shows the result from a least-squares bisector fit.
The AGNs were excluded from these 
fittings. Also shown in cyan and labelled individually are a few high-$z$ galaxies 
from Table~1, with $C(60/100)$ estimated from the published SEDs. 
The $T_{\rm dust}$ and $\Sigma_{\rm SFR}$ marks at the top of the plots
are explained in the text.
}
\label{Fig2}
\end{figure}

Many local dwarf galaxies with $Z \lesssim 0.5\,Z_{\odot}$ are relatively faint 
in CO\,(1$-$0), with a CO\,(1$-$0)/IR ratio being 1-2 orders of magnitude less than 
that for normal spirals (e.g., Schruba \etal 2012). This is usually 
attributed to a more severe CO dissociation by UV photons because of 
a lower dust opacity.  Nevertheless, the mid-$J$ CO line emission
arises from dense molecular clouds; possible UV self-shielding (Lee 
\etal 1996) implies that CO\,(7$-$6)/IR should not be as severely 
dependent on metallicity as CO\,(1-0)/IR.  
In Fig.~1, both Haro\,11 and He\,2-10 appear to have somewhat lower flux ratios
of CO\,(7$-$6) to IR or FIR. On the other hand, the bright \ion{H}{2} regions in
IC\,10 and NGC\,4214 are not much different from the (U)LIRGs in terms 
of the CO\,(7$-$6)/FIR ratio.  
Therefore, the low-metallicity dwarf galaxies examined 
here show only a moderately lower CO\,(7$-$6)/FIR ratio on average, but 
still within $\sim$2$\,\sigma_s$ of the average ratio for the (U)LIRGs.  


\subsection{{\rm \CII/CO\,(7$-$6)}, {\rm \NII/CO\,(7$-$6)} and $C(60/100)$} \label{sec3.2}

In Fig.~2 we plot the \CII/CO\,(7$-$6), \NII/CO\,(7$-$6) and \NII/\CII\ luminosity 
ratios as a function of $C(60/100)$ for the local (U)LIRGs.
All the plots span 2.3 dex vertically for direct comparison. The dust temperature, 
$T_{\rm dust}$, and $\Sigma_{\rm SFR}$ marks at the top of the plots were derived respectively 
from $C(60/100)$ assuming a dust emissivity index of $\beta = 1.5$ and from 
eq.~(2), which represents a least-squares bisector (Isobe et al.~1990) result
on a sample of 175 local star-forming galaxies [including 66 (U)LIRGs] in Liu 
et al.~(2015) with {\LIR}-based $\Sigma_{\rm SFR}$ and $0.25 \le C(60/100) \lesssim 1.1$. 
As in Chanial \etal (2007), both the normal galaxies and (U)LIRGs in this sample 
follow a single trend continuously. However, the r.m.s.~scatter at a given $C(60/100)$ 
is still significant, up to $\sim$0.9 dex. Moreover, eq.~(2) is calibrated up to 
$C(60/100) \sim 1.1$.
\begin{eqnarray}
\log\,\Sigma_{\rm SFR}/({\rm M}_{\odot}\,{\rm yr}^{-1}\,{\rm kpc}^{-2}) \hspace{1.5in} \nonumber \\
= (10.09 \pm 0.46)\,\log\,C(60/100) + (2.68 \pm 0.17). \ \ \ \ 
\end{eqnarray}

The data trends in Fig.~2 remain unchanged if \LIR\ or \LFIR\ replaces 
$L_{\rm CO(7-6)}$. All the three line ratios are anti-correlated with 
$C(60/100)$. 
Eqs.~(3) to (5) give the results from a vertical linear 
regression (i.e., the thin solid lines in Fig.~2) using the local (U)LIRG
detections in each plot (after excluding the AGNs, but including NGC\,6240 
that behaves ``normal'' here).  The resulting $\sigma_s$
with respect to the fit are 0.30, 0.23 and 0.15 dex, respectively.  
If we had further limited the GOALS galaxies
to a subset of 69 galaxies with $f_{70\mu{\rm m}}(17\arcsec) > 
85\%$, the resulting fit would be similar.
As demonstrated in Zhao \etal (2013), at a given $C(60/100)$, the scatter 
in the \NII/FIR [thus, \NII/CO\,(7$-$6)] flux ratios for SF-dominated galaxies
is largely driven by the hardness of the underlying radiation field. 
The reduced scatter in the \NII/\CII\ ratios suggests that a major part of 
the scatter in the \CII/CO\,(7$-$6) ratios should also be driven by 
the radiation hardness.  
\begin{eqnarray}
\log\,\CII/{\rm CO}\,(7{\rm -}6)  & = (-0.98 \pm 0.14)\,C(60/100) \nonumber \\
				  & \ \ \ \ + (2.47  \pm 0.11). \\ 
\log\,\NII/{\rm CO}\,(7{\rm -}6)  & = (-1.43 \pm 0.12)\,C(60/100) \nonumber \\
				  & \ \ \ \ + (1.69  \pm 0.09). \\ 
\log\,\NII/\CII   \ \ \ \ \ \             & = (-0.65 \pm 0.08)\,C(60/100) \nonumber \\
				  & \ \ \ \ - (0.66  \pm 0.06).
\end{eqnarray}


\tabletypesize{\scriptsize}
\begin{deluxetable*}{llccccccl}
\tablenum{1}
\tablewidth{0pt}
\tablecaption{High-redshift Galaxies}
\tablehead{
\colhead{Galaxy}   & \ \ $z$   & \colhead{Type}  & \colhead{CO\,(7$-$6)/FIR$^{a,b}$}    & \colhead{\CII/CO\,(7$-$6)$^a$}  & \colhead{\NII/CO\,(7$-$6)$^a$}  
        & \colhead{\NII/\CII$^a$}  & \colhead{$C(60/100)^c$}  & \colhead{References$^d$} \\
\colhead{ (1) } & \colhead{ (2) }  & \colhead{ (3) }  & \colhead{ (4) } & \colhead{ (5)} & \colhead{ (6)} & \colhead{ (7)} & \colhead{ (8)}  & \colhead{ (9)}}
\startdata
%
IRAS F10214+4724     & 2.286  &   QSO    &     -5.00(0.10)    & \nodata          & \nodata               & \nodata           &  \nodata      & (1,1,-,-,-)        \\  
SMM J213511-0102     & 2.32   &   SMG    &     -4.81(0.02)    & 2.20(0.08)       & \nodata               & \nodata           &  0.6          & (2,2,3,-,3)        \\  
SMM J16365+4057      & 2.383  &   SMG    &     -4.19(0.04)    & \nodata          & \nodata               & \nodata           &  \nodata      & (2,2,-,-,-)        \\  
SMM J16358+4105      & 2.452  &   SMG    &     -4.36(0.06)    & \nodata          & \nodata               & \nodata           &  \nodata      & (2,2,-,-,-)        \\  
SMM J04431+0210      & 2.509  &   SMG    &     -5.10(0.11)    & \nodata          & \nodata               & \nodata           &  \nodata      & (2,2,-,-,-)         \\  

Cloverleaf           & 2.558  &   QSO    &     -4.20(0.07)    & \nodata          & \nodata               & \nodata           &  \nodata      & (1,1,-,-,-)         \\  
SMM J14011+0252      & 2.565  &   SMG    &     -4.77(0.06)    & \nodata          & \nodata               & \nodata           &  \nodata      & (1,1,-,-,-)         \\  
VCV J1409+5628       & 2.583  &   QSO    &     -4.95(0.08)    & \nodata          & \nodata               & \nodata           &  \nodata      & (1,1,-,-,-)        \\  

AMS12                & 2.767  &   QSO    &     -4.99(0.05)    & \nodata          & \nodata               & \nodata           &  \nodata      & (2,2,-,-,-)        \\    
RX J0911+0551        & 2.796  &   QSO    &     -4.68(0.03)    & \nodata          & \nodata               & \nodata           &  \nodata      & (2,2,-,-,-)         \\  
HLSW-01              & 2.957  &   SMG    &     -4.52(0.05)    & 1.56(0.09)       & \nodata               & \nodata           &  1.3          & (2,2,4,-,4)         \\
MM18423+5938         & 3.930  &   SMG    &     -4.87(0.05)    & \nodata          & \ \ 0.53(0.02)        & \nodata           &  \nodata      & (2,2,-,2,-)         \\  
SMM J123711+6222     & 4.055  &   SMG    &     -4.71(0.05)    & \nodata          & \nodata               & \nodata           &  0.9          & (2,5,-,-,6)         \\ 
ID 141               & 4.243  &   SMG    &     -4.73(0.04)    & 1.59(0.10)       & \nodata               & \nodata           &  0.9          & (2,2,7,-,7)         \\  
BR 1202-0725 (N)     & 4.69   &   SMG    &     -4.12(0.05)    & 1.05(0.08)       & $<$0.50(n/a)          & $<$-0.44(n/a)     &  \nodata      & (2,8,9,10,-)         \\   
BR 1202-0725 (S)     & 4.69   &   QSO    &     -4.60(0.04)    & 1.00(0.07)       & $<$0.04(n/a)          & $<$-0.79(n/a)     &  \nodata      & (2,8,9,10,-)        \\   
LESS J033229-2756    & 4.755  &   SMG    &     \nodata        & \nodata          & \nodata               & \ \ -1.18(0.05)   &  0.9          & (-,-,11,2,12)        \\  
HDF850.1	     & 5.183  &   SMG    &     -4.74(0.06)    & 1.99(0.06)       & \nodata               & \nodata           & $\lesssim$1.0 & (2,13,14,-,14)      \\
HLS J091828+5142     & 5.243  &   SMG    &     -4.46(0.02)    & 1.38(0.02)       & \ \ 0.10(0.05)        & \ \ -1.28(0.05)   &  1.0          & (15,15,15,15,15)       \\  
HFLS3                & 6.34   &   SMG    &     -4.34(0.10)    & 1.07(0.15)       & $<$0.59(n/a)          & $<$-0.48(n/a)     &  1.2          & (16,16,16,16,16)       \\
SDSS J1148+5251      & 6.419  &   QSO    &     -4.61(0.05)    & 1.16(0.06)       & $<$0.13(n/a)          & $<$-1.03(n/a)     &  \nodata      & (2,2,17,17,-)              
\enddata
\tablenotetext{a}{Logarithmic luminosity ratios with the 1$\,\sigma$ uncertainty in the parentheses.}
\tablenotetext{b}{The quoted uncertainty does not include the (likely significant) error in \LFIR.}
\tablenotetext{c}{$C(60/100)$ estimated from the literature SED fit of a moderate to good quality.}
\tablenotetext{d}{Reference indices for the FIR, CO\,(7$-$6), \CII\ and \NII\ fluxes and $C(60/100)$, respectively, where the indices refer to
(1) see Solomon \& Vanden Bout (2005) for original reference,
(2) see Carilli \& Walter (2013) for the original reference,
(3) Ivison \etal (2010),
(4) Magdis \etal (2014),
(5) Carilli \etal (2010),
(6) Tan \etal (2014),
(7) Cox \etal (2011),
(8) Salom\'e \etal (2012),
(9) Carilli \etal (2013),
(10) Decarli \etal (2014a),
(11) De Breuck \etal (2014),
(12) Gilli \etal (2014),
(13) Decarli \etal (2014b),
(14) Walter \etal (2012),
(15) Rawle \etal (2014),
(16) Riechers \etal (2013), and 
(17) Walter \etal (2009).}
\end{deluxetable*}

A useful application of Fig.~2 at high $z$ is to infer $C(60/100)$ by 
measuring two of the three lines. Given its steeper 
correlation with $C(60/100)$ and modest sample scatter ($\sim$0.23 dex), 
the \NII/CO\,(7$-$6) ratio is the preferred one. While eq.~(4) is suitable
for inferring the \NII/CO\,(7$-$6) ratio from a measured $C(60/100)$, 
the reverse inference normally requires a regression of $C(60/100)$ on 
\NII/CO\,(7$-$6), of which the slope might be biased somewhat due to 
the selection effect that, at the low $C(60/100)$ end, our LIRG luminosity
cutoff may have left out some FIR-colder galaxies at a fixed 
\NII/CO\,(7$-$6) ratio. As a compromise, the thick solid line in Fig.~2b 
or eq.~(6) gives the least-squares bisector result 
as our favored estimator for $C(60/100)$.
\begin{eqnarray}
C(60/100) & = (-0.55 \pm 0.04)\,\log\,\NII/{\rm CO}\,(7{\rm -}6) \nonumber \\
          &  \ \ \ \ \  + (1.09 \pm 0.03).
\end{eqnarray}
The resulting scatter in $C(60/100)$ relative to the fit is $\sim$0.15,  
equivalent to an accuracy of $\lesssim$4\,K in $T_{\rm dust}$ in the case of 
a graybody emission with $T_{\rm dust} \gtrsim 30$\,K and a dust emissivity 
index $\beta \ge 1$.  However, it should be noted that eqs.~(3)-(5) are 
applicable only for $C(60/100) \gtrsim 0.4$,
below which the \NII/IR trend flattens out (Zhao et al.~2013).
Fig.~2 also hints that 
either a strong AGN or a low metallicity might enhance \NII/CO\,(7$-$6) 
and/or \CII/CO\,(7$-$6). While more data are needed to confirm these possible 
systematics, it has been known that a low metallicity tends to increase 
\CII/FIR (e.g., Madden \etal 1997)

\begin{figure}
\centering
\includegraphics[width=.55\textwidth, bb=50 144 592 690]{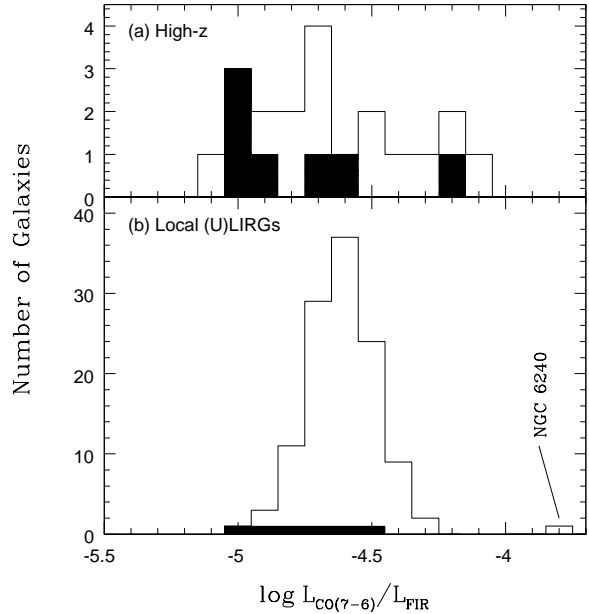}
\vspace{-0.7in}
\caption{
Histograms of $\log\,L_{\rm CO\,(7-6)}$/\LFIR, separately for (a) the high-$z$ 
sample and (b) our local (U)LIRG galaxies from Fig.~1b. In panel (a), the shaded 
part represents the QSOs; in (b), the 6 powerful AGNs are further shaded.}
\label{Fig3}
\end{figure}

\vspace{0.2in}
\section{Application to High-$z$ Galaxies} \label{sec4}

Table~1 lists the high-$z$ galaxies with either a CO\,(7$-$6) flux or both 
\CII\ and \NII\ observations in the literature. 
They consist of 14 sub-mm selected galaxies (SMGs) and 
7 quasars (QSOs). 
The line luminosities in solar units were derived using the formulae in 
Solomon \& Vanden Bout (2005).

There is a considerable uncertainty as to whether an IR luminosity given 
in the literature for a high-$z$ galaxy is \LIR\ or \LFIR.  When this 
distinction is not clear, the IR luminosity was usually either derived
from a FIR-radio correlation or scaled from a sub-mm flux density (see 
Carilli \& Walter 2013), and is therefore closer to \LFIR\ than \LIR.  
We therefore compare in Fig.~3 the CO\,(7$-$6)/FIR ratios between 
the local (U)LIRGs from Fig.~1 and the high-$z$ sample. Considering
the small sample size and the fact that the high-$z$ galaxies likely
have a much larger error in their \LFIR, we focus on the sample average.
The $\log L_{\rm CO\,(7-6)}$/\LFIR\ averages are ($-4.76 \pm 0.11$), 
($-4.59 \pm 0.08$) and ($-4.61 \pm 0.01$) for the high-$z$ QSOs, SMGs 
and the local (U)LIRGs (excluding NGC\,6240), respectively. 
A Student's t-test, allowing for unequal variances (Press 
et al.~1992, p.~617), showed an 80\% (23\%) confidence for the local 
(U)LIRGs and the high-$z$ SMGs (QSOs) to share the same average
CO\,(7$-$6)/FIR ratio.  The high-$z$ QSOs as a class have a   
lower average ratio than either the high-$z$ SMGs or 
the local (U)LIRGs. This is consistent with the sample selections, 
i.e., the QSOs should have stronger AGNs on average than the SMGs.
On the other hand, six out of the 7 high-$z$ QSOs possess the  
$L_{\rm CO\,(7-6)}$/\LFIR\ ratios comparable to those of the 6 local 
AGNs.  This also suggests that the high-$z$ galaxies with the lowest 
CO/FIR ratios are likely caused by the AGN rather than by a low metallicity. 

In Fig.~2 we also plotted those high-$z$ galaxies with available 
$C(60/100)$ in Table~1.  They all appear to be consistent with the trends 
defined by the local (U)LIRGs, with an overall agreement within 
$\sim$1$\,\sigma_s$ when only the detections are considered.  There are
3 galaxies in Table~1 with at least two of the 3 line ratios measured, 
but not $C(60/100)$.  In every case, the line ratios are consistent 
with each other within the context of Fig.~2.

Hodge \etal (2015) measured $\Sigma_{\rm SFR} \approx 119\,$M$_{\odot}$\,yr$^{-1}$\,kpc$^{-2}$
in J\,123711+6222, comparable to the predicted $\sim$165\,M$_{\odot}$\,yr$^{-1}$\,kpc$^{-2}$ 
from our eq.~(2).  For HFLS3 [$C(60/100) \sim 1.2$], the measured 
$\Sigma_{\rm SFR} \approx$ 0.73$-$1.15 $\,\times 10^3\,$M$_{\odot}$\,yr$^{-1}$\,kpc$^{-2}$ 
(Riechers \etal 2013). Eq.~(2) is calibrated only up to $C(60/100) \sim 1.1$, 
at which it predicts a mean $\Sigma_{\rm SFR} \approx 10^3\,$M$_{\odot}$\,yr$^{-1}$\,kpc$^{-2}$.

Our work here offers a simple, empirical method of using just two spectral
lines, CO\,(7$-$6) and \NIIline, to measure both SFR and $C(60/100)$ in 
(U)LIRGs.  Both lines suffer 
little dust extinction and are among the brightest FIR/sub-mm cooling 
lines, making this method particularly suited for probing the SF activity
in high-$z$ galaxies.  With a modern interferometric facility such as the Atacama 
Large Millimeter Array, both lines become observable at $z \gtrsim 0.5$. Thus, 
this technique enables 
the simultaneous study of the physical conditions (e.g., size) of the ionized
and dense molecular gas (e.g., Xu \etal 2015) as well as the SF activity 
across a wide redshift range.

\acknowledgments

This paper benefited from a number of thoughtful comments made by an 
anonymous referee.
Support for this work was provided in 
part by NASA through an award issued by JPL/Caltech. Y.Z. \& Y.G.
acknowledge support by NSFC grants No.~11390373 and 
11420101002, and CAS pilot-b program No.~XDB09000000.  
TDS acknowledges support by ALMA-CONICYT Grant No.~31130005.


%

\newpage


\begin{references}
\reference{} Armus, L., Mazzarella, J. M., Evans, A. S., et al.~2009, \pasp, 121, 559
\reference{} Asplund, M., Grevesse, N., Sauval, A. J., \& Scott, P.~2009, \araa, 47, 481
\reference{} Bendo, G. J., Galliano, F., \& Madden, S. C. 2012, \mnras, 423, 197
\reference{} Carilli, C. L., Daddi, E., Riechers, D., et al.~2010, \apj, 714, 1407
\reference{} Carilli, C. L., Riechers, D., Walter, F., Maiolino, R., Wagg, J., Lentati, L., McMahon, R., \& Wolfe, A. 2013, \apj, 763, 120
\reference{} Carilli, C. L., \& Walter, F.~2013, \araa, 51, 105
\reference{} Chanial, P., Flores, H., Guiderdoni, B, Elbaz, D., Hammer, F., \& Vigroux, L. 2007, \aap, 462, 81
\reference{} Cormier, D., Madden, S. C., Lebouteiller, V., et al.~2015, \aap\ (accepted) (arXiv:1502.03131)
\reference{} Cox, P., Krips, M., Neri, R., et al.~2011, \apj, 740, 63
\reference{} Daddi, E., Elbaz, D., Walter, F., et al. 2010, \apj, 714, L118
\reference{} Dale, D., Helou, G., Contursi, A., Silbermann, N., \& Kolhatkar, S.~2001, \apj, 549, 215
\reference{} De Breuck, C., Williams, R. J., Swinbank, M., et al.~2014, \aap, 565, A59
\reference{} Decarli, R., Walter, F., Carilli, C., et al.~2014a, \apj, 782, 17
\reference{} Decarli, R., Walter, F., Carilli, C., et al.~2014b, \apj, 782, 78
\reference{} D\'iaz-Santos, T., Armus, L., Charmandaris, V., et al. 2013, \apj, 774, 68
\reference{} Draine, B. T. \& Li, A. 2007, \apj, 657, 810
\reference{} Elbaz, D., Dickinson, M., Hwang, H. S., et al.~2011, \aap, 533, A119
\reference{} Farrah, D., Bernard-Salas, J., Spoon, H. W. W., et al. 2007, \apj, 667, 149
\reference{} Farrah, D., Lebouteiller, V., Spoon, H. W. W., et al. 2013, \apj, 776, 38
\reference{} Gilli, R., Norman, C., Vignali, C., et al.~2014, \aap, 562, 67
\reference{} Griffin, M. J., Abergel, A., Abreu, A., et al. 2010, \aap, 518, L3
\reference{} Helou, G., Soifer, B. T., Rowan-Robinson, M. 1985, \apj, 298, L7
\reference{} Hodge, J. A., Riechers, D., Decarli, R., Walter, F., Carilli, C. L., Daddi, E., \& Dannerbauer, H. 2015, \apj, 798, 18
\reference{} Inami, H., Armus, L., Charmandaris, V., et al.~2013, \apj, 777, 156
\reference{} Isobe, T., Feigelson, E. D., Akritas, M. G., \& Babu, G. J.~1990, \apj, 364, 104
\reference{} Ivison, R. J., Swinbank, A. M., Swinyard, B., et al.~2010, \aap, 518, L35
\reference{} Kennicutt, R. C., Jr.~1998, \araa, 36, 189
\reference{} Le Fl\'och, E., Papovich, C., Dole, H., et al.~2005, \apj, 632, 169
\reference{} Lee, H.-H., Herbst, E., Pineau des For\^ets, G., Roueff, E., \& Le Bourlot, J.~1996, \aap, 311, 690
\reference{} Liu, L., Gao, Y., \& Greve, T. R.~2015, \apj\ (accepted) (arXiv:1502.08001)
\reference{} Lu, N., Zhao, Y., Xu, C. K., et al. 2014, \apj, 787, L23 (Paper I)
\reference{} Madden, S. C., Poglitsch, A., Geis, N., Stacey, G. J., \& Townes, C. H.~1997, \apj, 483, 200
\reference{} Madden, S. C., R\'emy-Ruyer, A., Galametz, M., et al. 2013, \pasp, 125, 600
\reference{} Magdis, G. E., Rigopoulou, D., Hopwood, R., et al.~2014, \apj, 796, 63
\reference{} Poglitsch, A., Waelkens, C., Geis, N., et al.~2010, \aap, 518, L2
\reference{} Press, W. H., Teukolsky, S. A., Vetterling, W. T., \& Flannery, B. P.~1992, Numerical Recipes
             in C (2nd ed.; Cambridge, UK: University Press)
\reference{} Rawle, T. D., Egami, E., Bussmann, R. S., et al.~2014, \apj, 783, 59
\reference{} R\'emy-Ruyer, A., Madden, S. C., Galliano, F., et al.~2013, \aap, 557, A95
\reference{} Riechers, D. A., Bradford, C. M., Clements, D. L., et al.~2013, \nat, 496, 329 
\reference{} Rujopakarn, W., Rieke, G. H., Eisenstein, D. J., \& Juneau, S. 2011, \apj, 726, 93
\reference{} Sanders, D. B., \& Mirabel, I. F.~1996, \araa, 34, 749 
\reference{} Schruba, A., Leroy, A. K., Walter, F., et al.~2012, \aj, 143, 138
\reference{} Salom\'e, P., Gu\'elin, M., Downes, D., Cox, P., Guilloteau, S., Omont, A., Gavazzi, R., \& Neri, R.~2012, \aap, 545, 57
\reference{} Solomon, P. M., \& Vanden Bout, P. A. 2005, \araa, 43, 677
\reference{} Tan, Q., Daddi, E., Magdis, G., et al.~2014, \aap, 569, A98
\reference{} Walter, F., Wei{\ss}, A., Riechers, D. A., Carilli, C. L., Bertoldi, F., Cox, P., \& Menten, K. M.~2009, \apj, 691, L1
\reference{} Walter, F., Decarli, R., Carilli, C., et al.~2012, \nat, 486, 233 
\reference{} Wu, R., Polehampton, E. T., Etxaluze, M., et al.~2013, \aap, 556, A116
\reference{} Xu, C. K., Cao, C., Lu, N., et al.~2015, \apj, 799, 11
\reference{} Zhao, Y., Lu, N., Xu, C. K., et al.~2013, \apj, 765, L13
\end{references}
\end{document}